\documentclass[conference]{IEEEtran}
\IEEEoverridecommandlockouts

\usepackage{SKConferencePaperV0001}

\title{Graph Signal Processing:  Filter\\Design and Spectral Statistics}
\author{\IEEEauthorblockN{Stephen Kruzick and Jos\'{e} M. F. Moura}
\IEEEauthorblockA{Carnegie Mellon University, Department of Electrical Engineering \\
5000 Forbes Avenue, Pittsburgh, Pennsylvania 15213}\thanks{Stephen~Kruzick (skruzick@andrew.cmu.edu) and Dr. Jos\'{e}~M.~F.~Moura (moura@ece.cmu.edu) are members of the Department of Electrical and Computer Engineering at Carnegie Mellon University in Pittsburgh, PA, USA.  This work was supported by NSF grant \#CCF1513936.}
}

\begin{document}
\maketitle

\begin{abstract}
Graph signal processing analyzes signals supported on the nodes of a graph by defining the shift operator in terms of a matrix, such as the graph adjacency matrix or Laplacian matrix, related to the structure of the graph.  With respect to the graph shift operator, polynomial functions of the shift matrix perform filtering.  An application considered in this paper, convergence acceleration filters for distributed average consensus may be viewed as lowpass graph filters periodically applied to the states.  Design of graph filters depends on the shift matrix eigendecomposition.  Consequently, random graphs present a challenge as this information is often difficult to obtain.   Nevertheless, the  asymptotic behavior of the shift matrix empirical spectral distribution provides a substitute for suitable random matrix models.  This paper employs deterministic approximations for empirical spectral statistics from other works to propose optimization criteria for consensus acceleration filters, evaluating the results through simulation.
\end{abstract}


\section{Introduction}\label{Introduction}
Signal processing applications increasingly benefit from analysis methods that account for underlying structure in data, often modeled by a graph~\cite{ASan1,DShu1}.  Graph signal processing analyzes data in terms of this structure by representing signals as functions on the graph nodes and by defining the shift operator as a matrix that respects the graph structure, such as the adjacency matrix or the Laplacian matrix~\cite{ASan1,DShu1}.  Under these definitions, left multiplication of polynomials in the graph shift matrix with the graph signal vector performs filtering~\cite{ASan1}.  When the graph shift matrix is diagonalizable, decomposition of the signal vector in the basis of shift matrix eigenvectors provides an analogy to the Fourier transform, where the eigenvectors serve as pure frequency signal components~\cite{ASan3}.  Because the shift matrix eigenvalues relate to the total variation of the eigenvectors, they provide a notion of frequency that can be used to order the eigenvectors~\cite{ASan3}.  For a shift matrix $W_{\mathcal{G}}$, the response of a graph filter $p\left(W_{\mathcal{G}}\right)$ to an eigenvector $\mathbf{v}$ of $W_{\mathcal{G}}$ with corresponding eigenvalue $\lambda$ is $p\left(W_{\mathcal{G}}\right)\mathbf{v}=p\left(\lambda\right)\mathbf{v}$~\cite{ASan2}.  Thus, the eigenvalues of the shift matrix play a critical role in graph filter design.  Random graphs and matrices result in random eigenvalues, which complicates the filter design process.

This paper examines design of graph filters used to improve the convergence rate of the distributed average consensus algorithm for large scale random networks.  In distributed average consensus, the network nodes must reach agreement on the mean of data distributed among the nodes through an iterative algorithm only using local communications~\cite{ROlf1}, a problem relevant to applications such as sensor data fusion~\cite{LXia1}, flocking of multiagent systems~\cite{ROlf2}, and processor load balancing~\cite{GCyb1}.   In each iteration of the algorithm, the network nodes update a state variable, initially set to the node data, by computing a linear combination of neighbor node states.  Thus, distributed average consensus on a graph $\mathcal{G}$ is described by the dynamic system
\begin{equation}\label{ConsensusState}
\mathbf{x}_{n+1}=W_{\mathcal{G}}\mathbf{x}_n
\end{equation}
where $\mathbf{x}_n$ collects the node states at time $n$, $\mathbf{x}_0$ collects the initial node data, and the weight matrix $W_{\mathcal{G}}$ describes the state update,  which respects the graph structure $\mathcal{G}$~\cite{SKar2}.  If the weight matrix $W_{\mathcal{G}}$ satisfies
\begin{equation}\label{ConsCond}
W_{\mathcal{G}}\mathbf{1}=\mathbf{1}, \enskip {\boldsymbol \ell}^\top W_{\mathcal{G}}={\boldsymbol\ell}^\top,\enskip \rho\left(W_{\mathcal{G}}-J_{\boldsymbol\ell}\right)<1
\end{equation}
where $\rho$ is the spectral radius and $J_{\boldsymbol\ell}=\mathbf{1}\mathbf{\boldsymbol\ell}^\top/{\boldsymbol\ell}^\top\mathbf{1}$ is the weighted average consensus matrix, then $\mathbf{x}_n$ will asymptotically approach a weighted average of the initial data $J_{\boldsymbol\ell}\mathbf{x}_0=(\mathbf{\boldsymbol\ell}^\top\mathbf{x}_0/{\boldsymbol\ell}^\top\mathbf{1})\mathbf{1}$ at rate closely related to $\ln\rho\left(W_{\mathcal{G}}-J_{\boldsymbol\ell}\right)$~\cite{SKar2}.  Note that if $W_{\mathcal{G}}$ is a doubly stochastic matrix, $\boldsymbol\ell=\mathbf{1}$ so the unweighted average is produced.

Convergence rate acceleration for distributed average consensus may be accomplished by periodically applying a filter to previous node states.  For a filter of degree $d$, the modified algorithm performs the state update in~\eqref{ConsensusState} and additionally sets
\begin{equation}
\mathbf{x}_{n}:={\sum_{k=0}^{d}} a_k \mathbf{x}_{n-d+k},\quad n \equiv 0~(\textrm{mod}~d)
\end{equation}
on every $d$th iteration, where the filter coefficients form the polynomial $p\left(W_{\mathcal{G}}\right)={\sum}_{k=0}^{d}a_kW_{\mathcal{G}}^k$~\cite{EKok1}.  If the filter satisfies
\begin{equation}
p\left(W_{\mathcal{G}}\right)\mathbf{1}=\mathbf{1}, \enskip {\boldsymbol \ell}^\top p\left(W_{\mathcal{G}}\right)={\boldsymbol\ell}^\top,\enskip \rho\left(p\left(W_{\mathcal{G}}\right)-J_{\boldsymbol\ell}\right)<1
\end{equation}
the state converges to the weighted average consensus at rate closely related to $1/d\ln \rho\left(p\left(W_{\mathcal{G}}\right)-J_{\boldsymbol\ell}\right)$~\cite{EKok1}.  Thus, consensus acceleration filters should be designed to reduce $\rho\left(p\left(W_{\mathcal{G}}\right)-J_{\boldsymbol\ell}\right)$.  For known network topologies and weight matrices with $K$ distinct eigenvalues, finite time consensus filters use polynomials with zeros at the $K-1$ distinct eigenvalues $\lambda\neq 1$ to reach consensus in a finite number of iterations, representing an extreme example potentially requiring high filter degree~\cite{ASan4}.  For filters of lower fixed degree $1\leq d\leq K-1$ and known weight matrices,~\cite{EKok1} formulates a semidefinite program that yields the optimal solution.  For random switching network topologies,~\cite{EKok1} also attempts to design consensus acceleration filters by applying the semidefinite program to the mean weight matrix.  However, this can lead to suboptimal results or even divergence when the eigenvalues of the mean matrix do not sufficiently approximate the random weight matrix eigenvalues.  Therefore, consensus acceleration filter design should employ, when possible, a more complete understanding of the weight matrix spectral statistics.

Because the consensus weight matrix respects the graph structure and  distributed average consensus asymptotically produces a constant vector, consensus acceleration filters can be viewed as lowpass graph filters.  The methods presented in this paper combine deterministic approximations of the empirical eigenvalue distribution of large scale random matrices with linear programming for Chebyshev approximation to optimize the convergence rate for large scale constant (with respect to time iterations) random networks.  Related literature features contrasting approaches including filters based on Chebyshev polynomials of increasing degree~\cite{EMon1}, ARMA filter response specifications selected independently from the graph~\cite{ALou1}, spectral clustering based on the smallest and second largest eigenvalue modulus~\cite{SApe1}, and different asymptotic methods~\cite{FGam1}. In this paper, section~\ref{Background} describes background information from random matrix theory used to describe the weight matrix spectral statistics.  Section~\ref{Methods} proposes the filter optimization problem and discusses numerical practicalities.  Section~\ref{Simulations} supports the proposed design method with simulation results and also discusses choice of weight matrix.  Finally, Section~\ref{Conclusion} provides concluding remarks.  For an extended version of this paper, refer to~\cite{SKru3}.

\vspace{-1pt}
\section{Background: Random Matrix Theory}\label{Background}

Filter design for signal processing on graphs depends on knowledge of the graph shift matrix eigenvalues.  As previously noted, the value of the filter polynomial at each eigenvalue determines the response to the corresponding eigenvector.  Therefore, spectral information should inform design criteria for the filter response.  However, for contexts in which the graph is subject to stochastic influences, the corresponding shift matrix and associated eigenvalues also become random variables.  With few exceptions~\cite{CTra1}, the joint eigenvalue distribution typically proves analytically elusive.  Nevertheless, for some matrices of large size~\cite{EWig1,VGir1,ZBai1,RCou1,SVer1}, useful information can be obtained through the asymptotic behavior of the empirical eigenvalue distribution defined below.

For an $N \times N$ matrix $W_N$ with real eigenvalues $\lambda_1,\ldots,\lambda_N$, the empirical spectral distribution function, defined by
\begin{equation}
F_{W_N}\left(\lambda\right)=\frac{1}{N}\sum_{i=1}^{N}\chi\left(\lambda_i\leq \lambda\right)
\end{equation}
where $\chi$ is the indicator function, counts the number of eigenvalues on the interval $\left(-\infty,\lambda\right]$~\cite{RCou1}.  Likewise, the empirical spectral density function, defined by
\begin{equation}
f_{W_N}\left(\lambda\right)=\frac{1}{N}\sum_{i=1}^{N} \delta\left(\lambda-\lambda_i\right)
\end{equation}
where $\delta$ is the Dirac delta function, indicates eigenvalue locations~\cite{RCou1}.  While these are function valued random variables due to the random eigenvalues, the empirical spectral distribution and density may sometimes be approximated by deterministic functions for random matrices of large size.  For a family $W_N$ of random matrices of dimensions $N\times N$ parameterized by $N$, the sequence $F_{W_N}$ of empirical spectral distributions may approach a limiting spectral distribution $F_{lim}$.  Well known examples include the Wigner semicircular law~\cite{EWig1}, the Marchenko-Pastur Law~\cite{RCou1}, and the Girko circular law~\cite{VGir1}.  Similarly, a sequence of deterministic distribution functions $F_N$ that asymptotically approximate the empirical spectral distribution sequence is known as a deterministic equivalent for the sequence~\cite{RCou1}.

The stochastic canonical equation methods of Girko provide one approach to obtaining these deterministic equivalents~\cite{VGir1}.  For random matrix models that  satisfy certain regularity conditions, the Stieltjes transform of a deterministic equivalent distribution can be computed by solving a system of equations that depends on the random matrix model parameters~\cite{VGir1}.  These methods allow analysis of matrices with elements that are independent but not necessarily identically distributed, as would arise from many random graph models, such as percolations of non-complete supergraphs~\cite{VGir1}.  Use of Girko's methods to approximate empirical spectral distributions of graph adjacency matrices, normalized adjacency matrices, and normalized Laplacian matrices has been examined in~\cite{KAvr1,SKru1,SKru2}.  This paper employs the deterministic equivalents computed using these methods along with the filter design criteria proposed in Section~\ref{Methods} to produce the simulation results appearing in Section~\ref{Simulations}.

\begin{figure}[t]
\includegraphics[width=\linewidth]{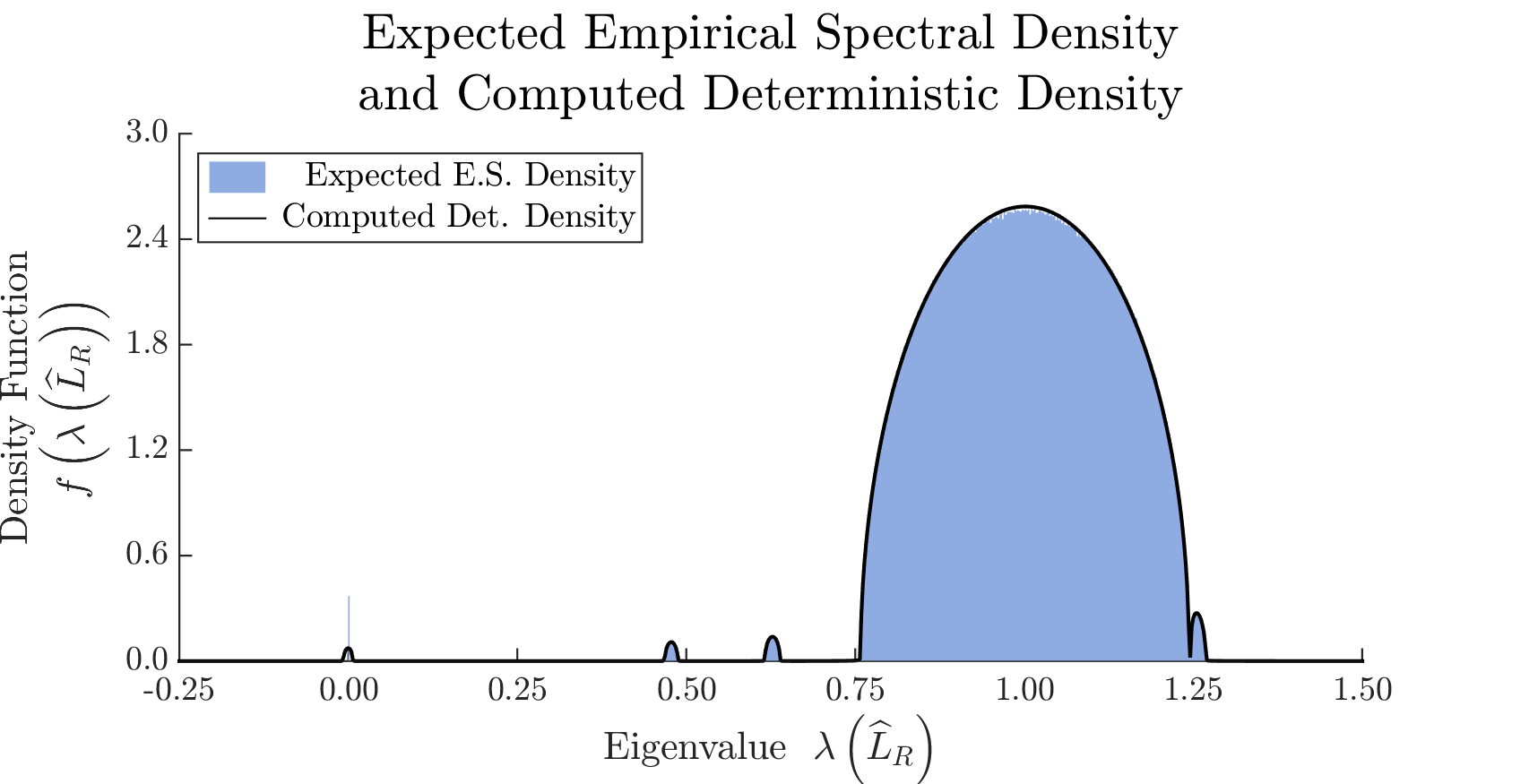}
\caption{Example empirical spectral distribution (blue shaded) and deterministic approximation (black curve) for row normalized Laplacian of $2\mathcal{D}$ lattice stochastic block model network from Figure~\ref{SimFigs_OptComp_Lat2SBM} described in Section~\ref{Simulations}}\label{ESDEx}
\end{figure}

\vspace{-1pt}
\section{Filter Design Method}\label{Methods}
In order to design graph filters that accelerate the distributed average consensus process, note that the worst case consensus convergence error at time  $n=md$ is $\left\|p\left(W\right)^m-J_{\boldsymbol\ell}\right\|_2$ where $d$ is the degree of filter polynomial $p\left(W\right)=\sum_{k=0}^{d}a_kW^k$ and $\boldsymbol\ell^\top W=\boldsymbol\ell^\top$.  When the weight matrix can be diagonalized by matrix $V$, it follows that 
\begin{equation}
\begin{gathered}
\rho\left(p\left(W\right)-J_{\boldsymbol\ell}\right)^m\leq\left\|p\left(W\right)^m-J_{\boldsymbol\ell}\right\|_2\\\
\leq \left\|V\right\|_2\left\|V^{-1}\right\|_2\rho\left(p\left(W\right)-J_{\boldsymbol\ell}\right)^m.
\end{gathered}
\end{equation}
Thus, bounds for the worst case convergence rate can be optimized by designing the filter to minimize $\rho\left(p\left(W\right)-J_{\boldsymbol\ell}\right)$.  When comparing performance of different length filters, the per iteration convergence rate $1/d\ln \rho\left(p\left(W\right)-J_{\boldsymbol\ell}\right)$ should be used.  Given a random $N\times N$ weight matrix model with deterministic approximation $f_N$ for the empirical spectral density, the proposed optimization~\eqref{ProposedOpt1} solves this problem in the space of polynomials with degree at most $d$.  The condition that $p\left(W\right)\mathbf{1}=\mathbf{1}$ and the condition that $W\mathbf{1}=\mathbf{1}$ impose the equality constraint $p(1)=1$.
\begin{equation}\label{ProposedOpt1}
\begin{gathered}
\begin{aligned}
\min_{p\in P_d} \max_{\lambda\in \Lambda_{\kappa,\tau}} &\left|p\left(\lambda\right)\right|\\
\mathrlap{\st}\hphantom{\min_{p\in P_d}\max_{\lambda\in \Lambda_{\kappa,\tau}}} &\hphantom{|}p\left(1\right)=1
\end{aligned}\\
\Lambda_{\kappa,\tau}=\left\{\lambda<1-\kappa|f_N\left(\lambda\right)>\tau\right\}
\end{gathered}
\end{equation}
The set $\Lambda_{\kappa,\tau}$ for small constants $\kappa$ and $\tau$ defines the spectral region of interest, where $\kappa$ guarantees a transition region around the equality constraint and $\tau$ specifies where the density function $f_N$ has negligible value.  Intuitively,~\eqref{ProposedOpt1} minimizes the worst case graph filter frequency response at eigenvalues of $W$ captured as the support of the deterministic approximation $f_N$ to the empirical eigenvalue distribution.  While a loss of robustness is possible if an eigenvalue falls outside the approximate spectral distribution support, the true support is well approximated asymptotically large networks with spectral convergence behavior and reasonable filter degrees, and additional constraints could be added to explicitly prevent this.

This minimax polynomial optimization can be understood in the context of Chebyshev approximation and produces equiripple behavior.  While a Remez algorithm could be employed, it can also be formulated as a linear program for simplicity.  In practice, the substitution in~\eqref{EqP}-\eqref{EqQ} may be utilized to improve the conditioning of the resulting linear program by eliminating the equality constraint, where the $\left\{\phi_n\right\}$ denote a basis of scaled and shifted Chebyshev polynomials of the first kind chosen as in~\eqref{ChebyFirstKind}.
\begin{align}
p\left(\lambda\right)&=1+(1-\lambda) q\left(\lambda\right) \label{EqP}\\
q\left(\lambda\right)&=\sum_{n=0}^{d-1} a_n\phi_n\left(\lambda\right)
\label{EqQ}\end{align}
This results in the modified optimization problem in~\eqref{ProposedOpt3}, which can be used to obtain $p\left(\lambda\right)$ by finding $q\left(\lambda\right)$.
\begin{equation}\label{ProposedOpt3}
\begin{gathered}
\min_{ q\in P_{d-1}}{ \max_{\lambda \in \Lambda_{\kappa,\tau}} \left|(1-\lambda)\left(q\left(\lambda\right)+1/{(1-\lambda)}\right)\right|}\\
\Lambda_{\kappa,\tau}=\left\{\lambda<1-\kappa|f_N\left(\lambda\right)>\tau\right\}
\end{gathered}
\end{equation}
The linear program in~\eqref{GeneralSolutionLP} solves the problem by minimizing the bound on the objective function absolute value at representative sample points $\Lambda_S\subset \Lambda_{\kappa,\tau}$ (e.g., several hundred uniformly spaced points).
\begingroup
\thinmuskip=\muexpr\thinmuskip*1/8\relax
\medmuskip=\muexpr\medmuskip*1/8\relax
\begin{equation}
\begin{aligned}
\min_{\{a_n\},\epsilon>0}  & \hphantom{{}-{}}\epsilon\\[-3\jot]
\st 
&\hphantom{{}-{}}(1-\lambda_i)\left(\sum_{n=0}^{d-1} a_n \phi_n\left(\lambda_i\right)-\frac{1}{1-\lambda_i}\right)<\epsilon\\
&-(1-\lambda_i) \left(\sum_{n=0}^{d-1} a_n \phi_n\left(\lambda_i\right)-\frac{1}{1-\lambda_i}\right)<\epsilon\\
&\hphantom{{}-{}(1-\lambda_i) g\left(\lambda_i\right)\left(\sum_{n=0}^{d-1} a_n \phi_n\left(\lambda_i\right)-\frac{1}{1-\lambda_i}\right)<\epsilon}\mathllap{\textrm{for all }\lambda_i\in \Lambda_S}
\end{aligned}\label{GeneralSolutionLP}
\end{equation}
\endgroup
For good numerical performance, the basis of Chebyshev polynomials $\left\{\phi_n\right\}$ should be scaled to the spectral region of interest $\Lambda_{\kappa,\tau}$ as follows, where $T_n$ is the degree $n$ standard Chebyshev polynomial of the first kind~\cite{JBoy1}.
\begin{equation}\label{ChebyFirstKind}
\begin{gathered}
\phi_n\left(\lambda\right)=T_n\left(\frac{\lambda-\beta}{\alpha}\right)\\
\alpha=\frac{\lambda_{\max}-\lambda_{\min}}{2},\quad \beta=\frac{\lambda_{\max}+\lambda_{\min}}{2}\\
\lambda_{\max}=\max\left(\Lambda_{\kappa,\tau}\right), \quad \lambda_{\min}=\min\left(\Lambda_{\kappa,\tau}\right)
\end{gathered}
\end{equation}

\section{Simulations}\label{Simulations}
\floatsetup[subfloat]{floatrowsep=floathfill}

\begin{figure*}[t]
\ffigbox[\textwidth]
{

\begin{floatrow}[2]

\ffigbox[\linewidth]
{\includegraphics[width=\linewidth]{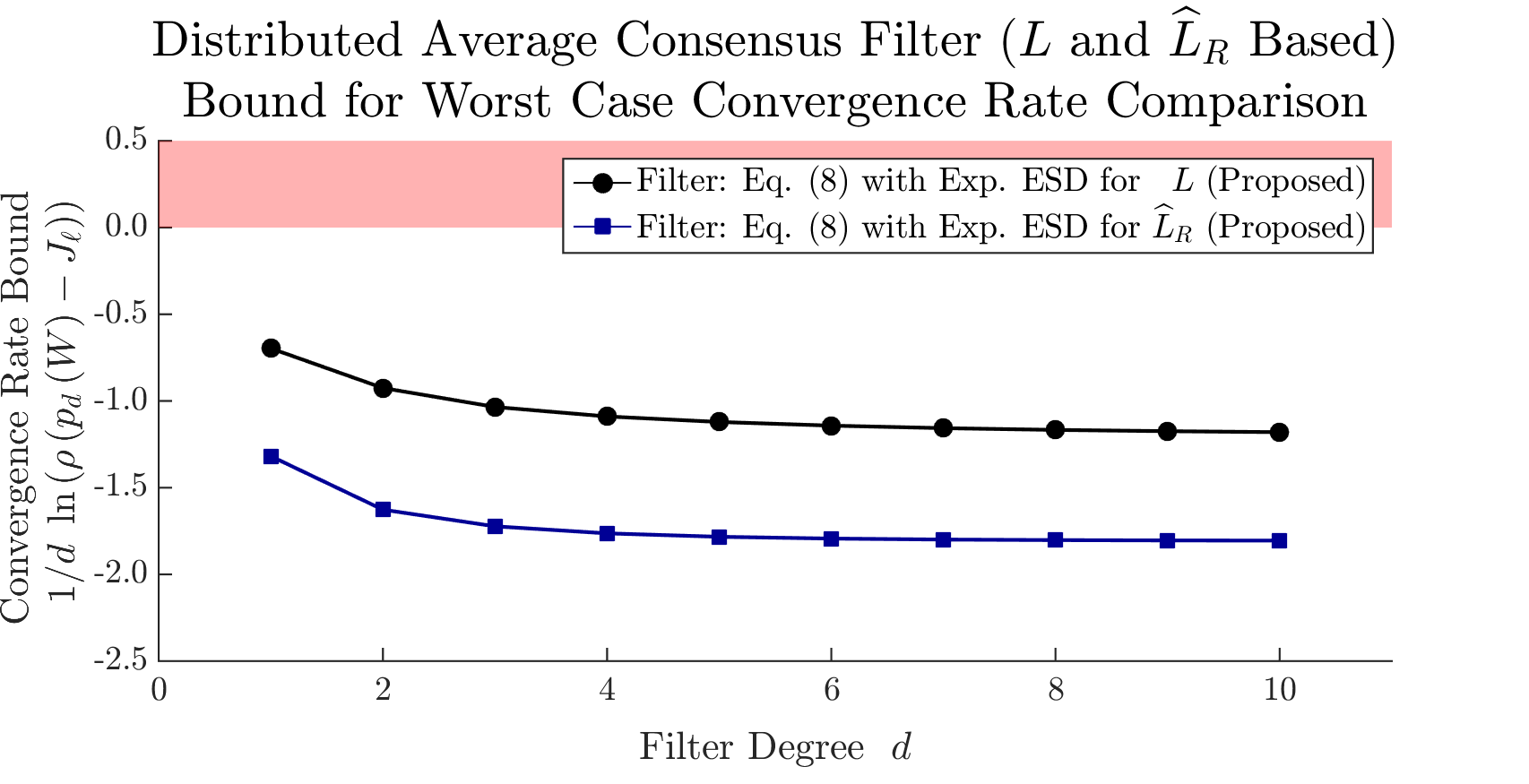}}
{\caption{Consensus convergence rates for an Erd\H{o}s-R\'{e}nyi network with $2000$ nodes and connection probability $\theta=0.03$ using filters based on weights $W\!\!=\!I\!-\alpha L$ and $W\!\!=\!I\!-\alpha \widehat{L}_R$}\label{SimFigs_WMatComp_ER}}
\hfill
\ffigbox[\linewidth]
{\includegraphics[width=\linewidth]{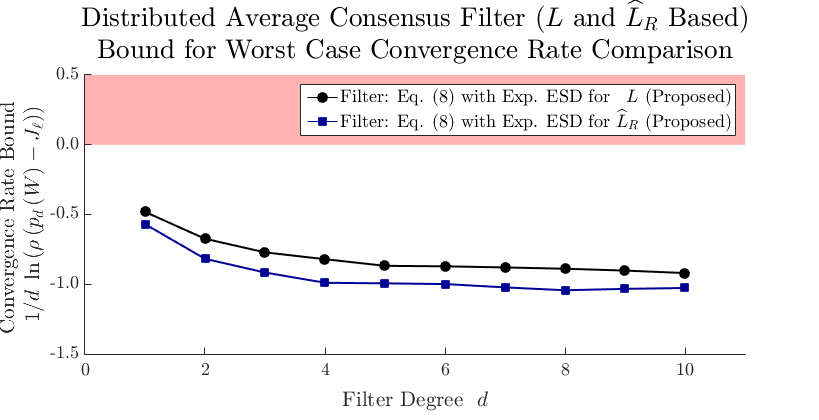}}
{\caption{Consensus convergence rates for a 2$\mathcal{D}$ lattice stochastic block model network with $3 \times 7$ populations of $100$ nodes and connection probabilities $(\theta_0,\theta_1,\theta_2)=(0.15,0.09,0.06)$ using filters based on weights $W\!\!=\!I\!-\alpha L$ and $W\!\!=\!I\!-\alpha \widehat{L}_R$}\label{SimFigs_WMatComp_Lat2SBM}}

\end{floatrow}

}{}
\vspace{.9\textfloatsep}
\ffigbox[\textwidth]
{

\begin{floatrow}[2]

\ffigbox[\linewidth]
{\includegraphics[width=\linewidth]{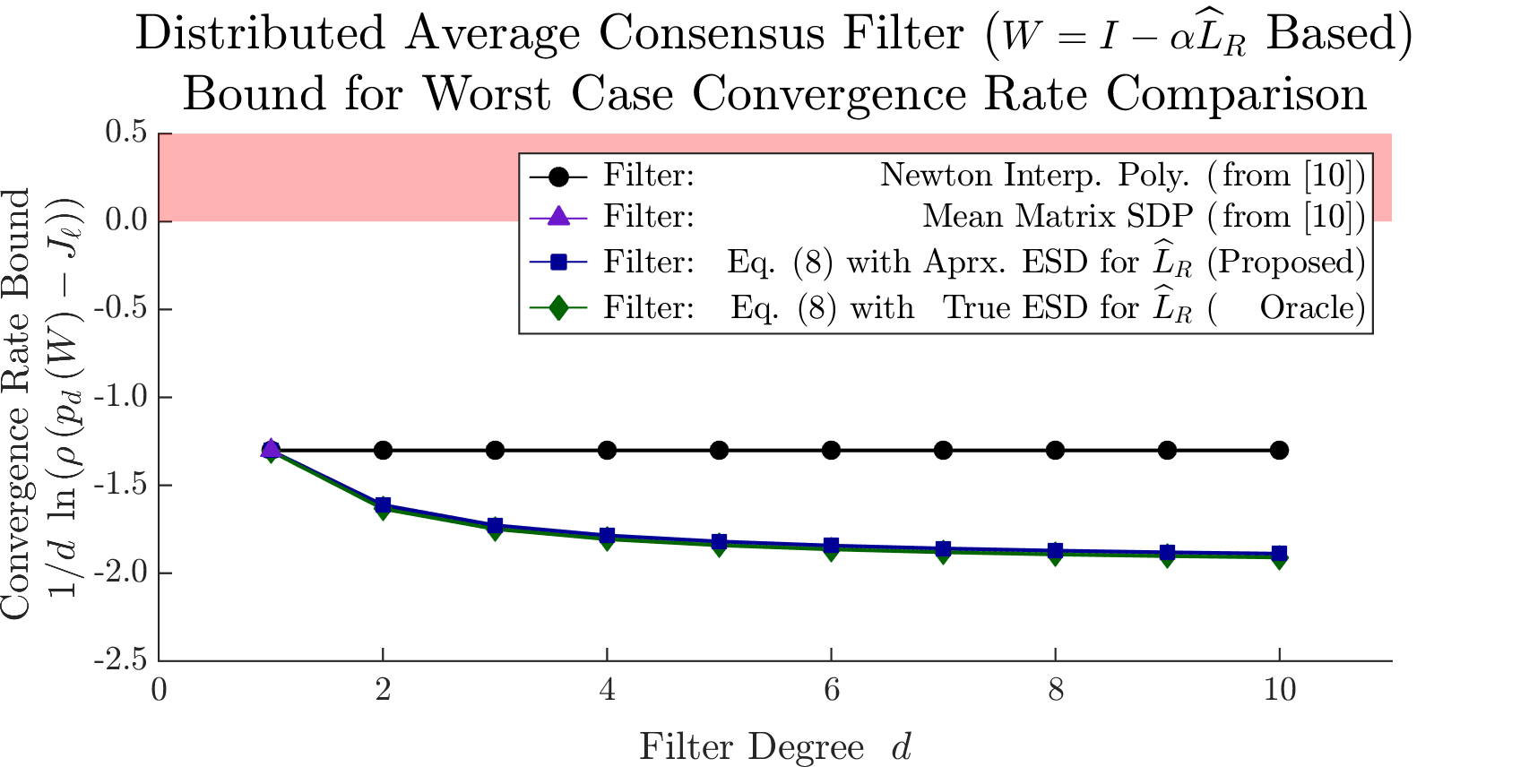}}
{\caption{Consensus convergence rates for an Erd\H{o}s-R\'{e}nyi network with $1000$ nodes and connection probability $\theta=0.05$ using filters of degree $d=1,\ldots,10$ based on $\widehat{L}_R$ for  listed methods}\label{SimFigs_OptComp_ER}}
\hfill
\ffigbox[\linewidth]
{\includegraphics[width=\linewidth]{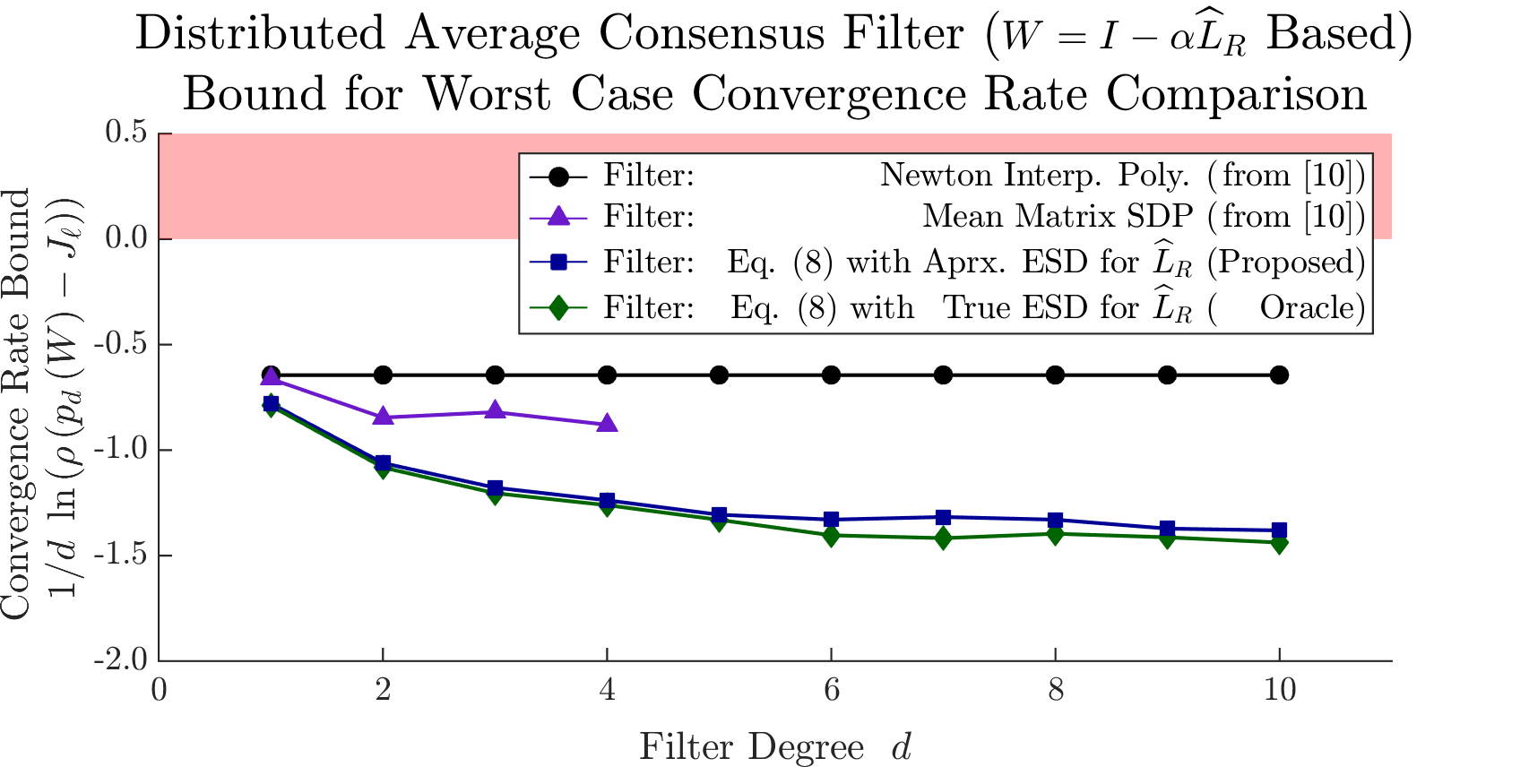}}
{\caption{Consensus convergence rates for a 2$\mathcal{D}$ lattice stochastic block model network with $3 \times 4$ populations of $100$ nodes and connection probabilities $(\theta_0,\theta_1,\theta_2)=(0.10,0.10,0.10)$ using filters of degree $d=1,\ldots,10$ based on $\widehat{L}_R$ for listed methods}\label{SimFigs_OptComp_Lat2SBM}}

\end{floatrow}

}{\vspace{15pt}}
\end{figure*}


\newlength{\declen}\setlength{\declen}{0pt}

In order to demonstrate the performance of the filters designed using the proposed methods, this section provides supporting simulation results for constant random networks.  The first pair of simulations compares the relative convergence rates of filters for weight matrices based on the unnormalized Laplacian matrix and weight matrices based on the row-normalized Laplacian matrix.  The second pair of simulations compares the proposed method to other filter design methods available in the literature.

\vspace{-\declen}
The simulations presented in this section cover two random network models.  An Erd\H{o}s-R\'{e}nyi network on $N$ nodes describes a random graph model in which each pair of nodes connects according to independent Bernoulli trials with link probability $\theta$~\cite{BBol1}.  A $\mathcal{D}$-dimensional lattice stochastic block model consists of $N_1\times\cdots\times N_\mathcal{D}$ populations, each with $M$ nodes.  The populations correspond to $\mathcal{D}$-tuples and collectively form a lattice in the sense of~\cite{RLas1}.  Nodes connect to other nodes according to independent Bernoulli trials if their population tuples differ by at most one symbol, with link probability $\theta_0$ within populations and $\theta_k$ between nodes in populations along lattice dimension $k$.  The adjacency matrix empirical spectral distributions of both of these models are amenable to analysis by the methods of Girko, as done in~\cite{KAvr1,SKru1,SKru2}.  The second simulation uses the spectral distribution approximation results from~\cite{KAvr1,SKru1,SKru2} without repeating the derivation.

\vspace{-\declen}
The proposed optimization problem does not specify a particular scheme for choosing the weight matrix $W_\mathcal{G}$ for graph $\mathcal{G}$, requiring only that $W_{\mathcal{G}}$ satisfy the weighted consensus conditions~\eqref{ConsCond}.  A common choice for the weight matrix, $W_\mathcal{G}=I-\alpha L\left(\mathcal{G}\right)$ depends on the unnormalized graph Laplacian matrix $L\left(\mathcal{G}\right)=D\left(\mathcal{G}\right)-A\left(\mathcal{G}\right)$ where $D\left(\mathcal{G}\right)$ is the diagonal degree matrix and $A\left(\mathcal{G}\right)$ is the graph adjacency matrix.  The scale parameter $\alpha$ must be chosen to satisfy the spectral radius constraint.  Note that $W_\mathcal{G}=I-\alpha L\left(\mathcal{G}\right)$ is a doubly stochastic matrix, so the unweighted average is produced.  However, the spectral statistics of the Laplacian are typically not approachable using Girko's methods.  In the case of Erd\H{o}s-R\'{e}nyi networks, the limit of the Laplacian empirical spectral distribution has density given by the free convolution \cite{SVer1} of a Gaussian distribution and a semicircular distribution~\cite{XDin1}, but results for other models are typically inaccessible.

\vspace{-\declen}
Alternatively, the weight matrix scheme $W_\mathcal{G}=I-\alpha \widehat{L}_R\left(\mathcal{G}\right)$ where $\widehat{L}_R\left(\mathcal{G}\right)=I-D\left(\mathcal{G}\right)^{-1}A\left(\mathcal{G}\right)$ is the row-normalized Laplacian can be selected.  
This is a row-stochastic matrix with left eigenvector $\boldsymbol\ell=\mathbf{d}$, the vector of node degrees, corresponding to $\lambda=1$.  Although this leads to a weighted average, the unweighted average can be produced through premultiplication by the corrective transform $\left(\mathbf{d}^\top\mathbf{1}\right)/\left(\mathbf{1}^\top\mathbf{1}\right)D^{-1}$, which can be applied at each node using the average node degree and the local node degree.  Because the row-normalized Laplacian is more amenable to analysis using the methods of Girko~\cite{KAvr1,SKru1,SKru2}, $W_\mathcal{G}=I-\alpha\widehat{L}_R\left(\mathcal{G}\right)$ presents an appealing choice for use with~\eqref{ProposedOpt1}.  
Figures~\ref{SimFigs_WMatComp_ER} and~\ref{SimFigs_WMatComp_Lat2SBM} show the relative convergence rates for each of these weight matrices using the simulated expected empirical densities for an Erd\H{o}s-R\'{e}nyi network and a $2\mathcal{D}$ lattice stochastic block model with $1/\alpha$ chosen as the approximate center of the distribution support.  Note that $W_\mathcal{G}=I-\alpha\widehat{L}_R\left(\mathcal{G}\right)$ outperforms $W_\mathcal{G}=I-\alpha L\left(\mathcal{G}\right)$ further recommending use of the weight matrix based on the row-normalized Laplacian.  Intuitively, this observation is expected due to the convolution-like composition of the Laplacian empirical spectral density resulting in less compact support.

\vspace{-\declen}
The second group of simulations compares the results obtained from the proposed design method~\eqref{ProposedOpt1} to those from the mean matrix semidefinite program method described in~\cite{EKok1} and from the Newton interpolating polynomial method described in~\cite{EKok1} for various filter degrees.  A deterministic approximation for the empirical spectral density was computed using Girko's stochastic canonical equation theorem as in~\mbox{\cite{KAvr1,SKru1,SKru2}} to be used with the proposed optimization method~\eqref{ProposedOpt1}.  The results for the proposed optimization method compare favorably to these methods as seen in Figures~\ref{SimFigs_OptComp_ER} and~\ref{SimFigs_OptComp_Lat2SBM}, which plot the per iteration convergence rates for filters of degree $d=1,\ldots,10$.  Note that because the polynomial based on the semidefinite program only has a unique solution for filter degree less than the number of distinct mean weight matrix eigenvalues, it only appears for $d=1$ in Figure \ref{SimFigs_OptComp_ER} and $d\leq 4$ in Figure \ref{SimFigs_OptComp_Lat2SBM}.  Furthermore, the results were compared with filters designed using the proposed optimization method~\eqref{ProposedOpt1} with an oracle for the true empirical spectral distribution.  The proposed method using the deterministic equivalent distribution achieves nearly equal results, demonstrating good performance.

\section{Conclusion}\label{Conclusion}
Filter design for signal processing on random graphs requires the ability to obtain information about the random shift matrix eigenvalues.  Thus, methods from random matrix theory that capture asymptotic deterministic structure in the empirical spectral distributions of suitable matrices provide useful tools for filter design in the context of large scale random graphs.  This paper proposed an optimization problem to derive convergence acceleration filters for distributed average consensus, which can be understood as lowpass graph filters.  In practical terms, these filters enable improved accuracy over a fixed number of iterations or a given level of accuracy in fewer iterations.  The proposed method combines Chebyshev approximation techniques with deterministic equivalents for shift matrices derived in other papers using Girko's stochastic canonical equations.  Simulation results demonstrate that the filters derived perform well on constant random networks, comparing favorably to other tested methods.  Consideration of weight matrices based on the unnormalized and row-normalized Laplacians suggest faster convergence can be achieved using random row-normalized Laplacians.  Further work will focus on analysis of time-varying random networks and additional graph signal processing applications.

\begingroup
\linespread{.96}
\bibliographystyle{IEEEbib}
\bibliography{GSPFilterDesign}
\endgroup

\end{document}